\documentclass [11pt, letterpaper]{article}
\pdfoutput=1 

\usepackage{fullpage}
\usepackage{amsmath}
\usepackage{amssymb}
\usepackage{graphicx}
\usepackage{mathrsfs}
\usepackage{wrapfig}
\usepackage{boxedminipage}
\usepackage{epsfig}
\usepackage{setspace}
\usepackage{subfigure}
\usepackage{float}
\usepackage{hyperref}
\usepackage{enumerate}
\usepackage{cite}
\usepackage[font=footnotesize,labelfont=rm]{caption}

\begin{document}

\begin{flushright}
{\tt arXiv:1602.02838}
\end{flushright}

{\flushleft\vskip-1.35cm\vbox{\includegraphics[width=1.25in]{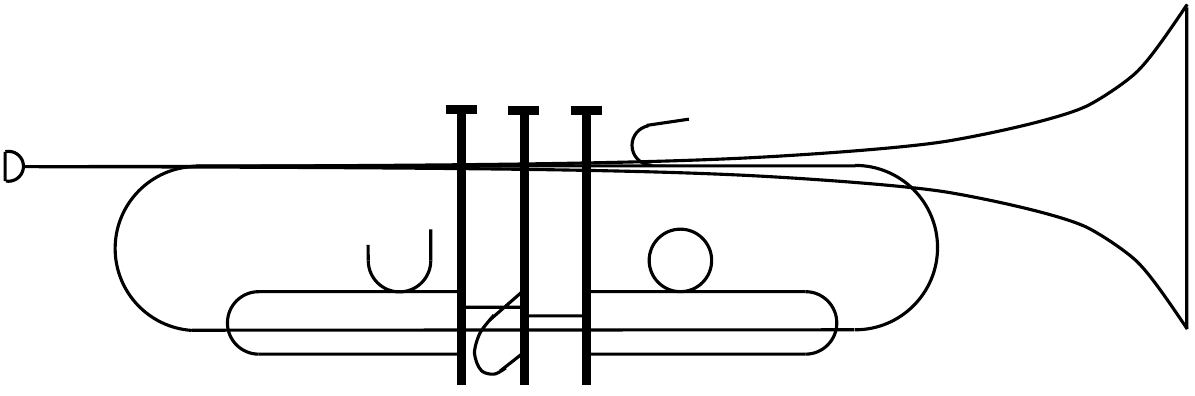}}}

\bigskip
\bigskip
\bigskip
\bigskip

\bigskip
\bigskip
\bigskip
\bigskip

\begin{center} 

{\Large\bf  An Exact Efficiency Formula  for Holographic Heat Engines}




\end{center}

\bigskip \bigskip \bigskip\bigskip

\centerline{\bf Clifford V. Johnson}

\bigskip
\bigskip
\bigskip
\bigskip

  \centerline{\it Department of Physics and Astronomy }
\centerline{\it University of
Southern California}
\centerline{\it Los Angeles, CA 90089-0484, U.S.A.}

\bigskip

\centerline{\small \tt johnson1,  [at] usc.edu}

\bigskip
\bigskip


\begin{abstract} 
\noindent  Further consideration is given to the efficiency of black hole heat engines that perform mechanical work {\it via} the $pdV$ terms   present in the First Law of extended gravitational thermodynamics. It is noted that when the engine cycle is a rectangle with sides parallel to the $(p,V)$ axes, the efficiency can be written simply in terms of the   mass of the black hole  evaluated at the corners. Since an arbitrary cycle can be approximated to any desired accuracy by a tiling of  rectangles, a  general geometrical algorithm for computing the efficiency follows. A simple generalization of the  algorithm renders it  applicable to more general classes of heat engine, beyond the black hole context.
\end{abstract}
\newpage \baselineskip=18pt \setcounter{footnote}{0}

\section{Introduction}
\label{sec:introduction}
This paper concerns the efficiency of holographic heat engines, which  were defined in ref.\cite{Johnson:2014yja}.  They are a natural concept in {\it extended}  gravitational thermodynamics, which, in making dynamical the cosmological constant ($\Lambda$)  in a theory of gravity, supplies\footnote{Here we are using geometrical units where $G,c,\hbar,k_{\rm B}$ have been set to unity. 
} a pressure variable $p=-\Lambda/8\pi$ and its conjugate volume~$V$ (see  refs.\cite{Kastor:2009wy,Caldarelli:1999xj,Wang:2006eb,Sekiwa:2006qj,LarranagaRubio:2007ut,Dolan:2010ha,Cvetic:2010jb,Dolan:2011jm,Dolan:2011xt,Henneaux:1984ji,Teitelboim:1985dp,Henneaux:1989zc}). 

 \begin{wrapfigure}{l}{0.3\textwidth}
{\centering
\includegraphics[width=1.8in]{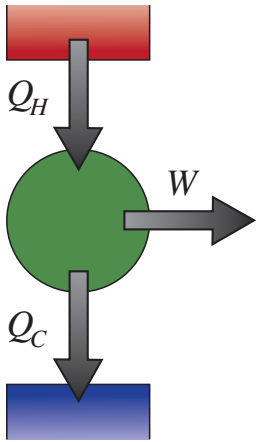} 
   \caption{\footnotesize  The flows in a cycle.}   \label{fig:cyclesa}
}
\end{wrapfigure}
One may extract mechanical work {\it via} the  $pdV$ term in the First Law of Thermodynamics, and so it is possible to define a cycle in state space during which there is a net input heat $Q_H$ flow, a net output heat flow $Q_C$, and a net output work $W$  such that  $Q_H=W+Q_C$. See figure~\ref{fig:cyclesa}. The efficiency is then $\eta=W/Q_H=1-Q_C/Q_H$. Its value is determined by the equation of state of the system  and the choice of cycle in state space. The gravitational solution (a black hole, in the cases studied here) supplies the equation of state: The temperature $T$, entropy $S$, and other quantities can be computed\cite{Bekenstein:1973ur,Bekenstein:1974ax,Hawking:1974sw,Hawking:1976de}, and there are relations between them. There's also a relation between the thermodynamic volume $V$ and the horizon radius of the black hole\cite{Kastor:2009wy}. The precise form of all these relations depends upon the type of black hole, and of course the parent theory of gravity under discussion. For example, refs.\cite{Johnson:2015ekr,Johnson:2015fva} study the  efficiency in the situation  when the parent gravity theory has  Gauss--Bonnet and Born--Infeld sectors.

 \begin{wrapfigure}{l}{0.3\textwidth}
{\centering
\includegraphics[width=1.8in]{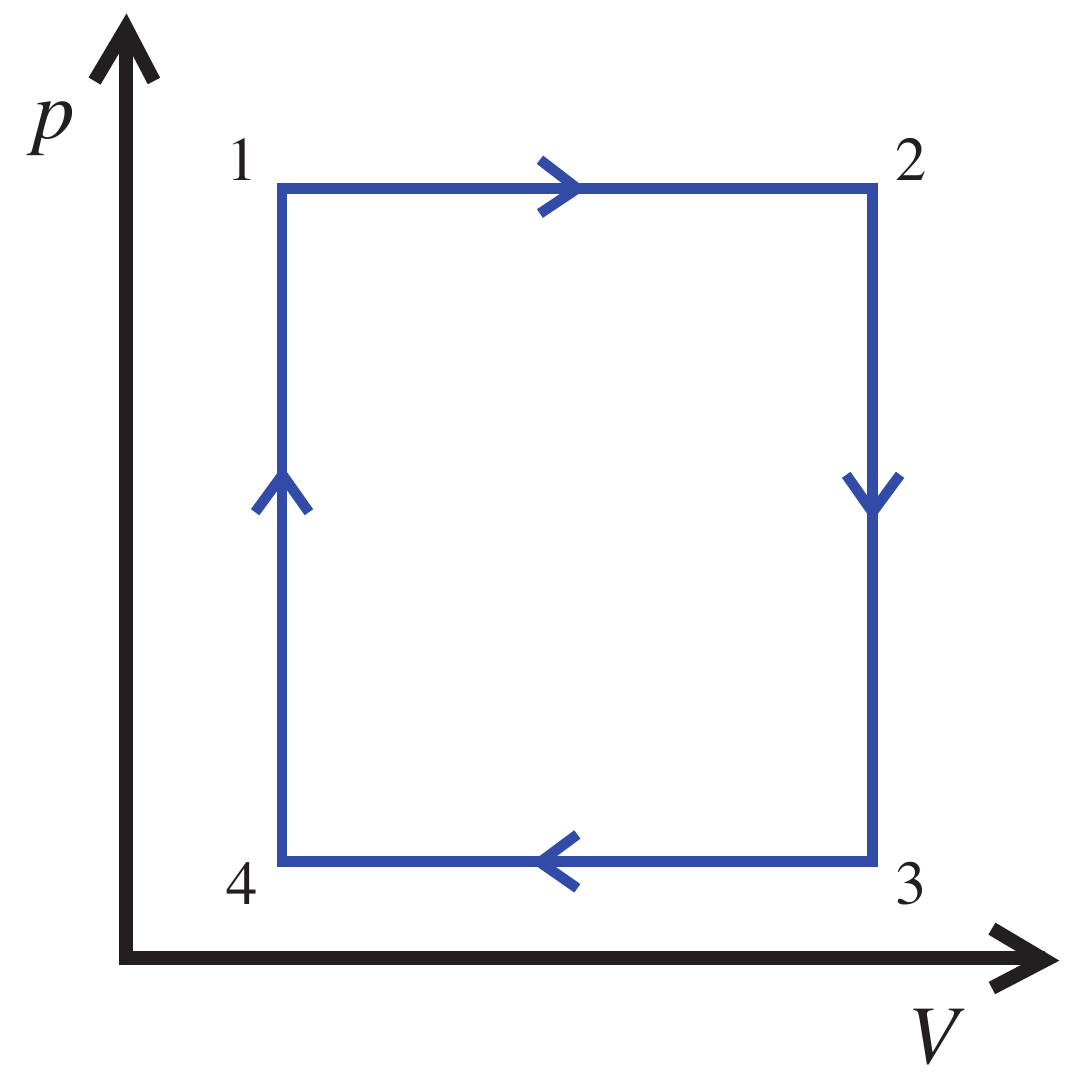} 
   \caption{\footnotesize  A special cycle.}   \label{fig:cyclesb}
}
\end{wrapfigure}
In this extended thermodynamics context,  we  work with a negative cosmological constant (defining a positive pressure), for which such physics has an holographic duality\cite{Maldacena:1997re,Witten:1998qj,Gubser:1998bc,Witten:1998zw,Aharony:1999t} to non--gravitational field theories in one dimension fewer, at large $N$ (where $N$ is the rank of a field theory gauge group, or an analogue thereof). 
This is why our heat engines in this context are called ``holographic''. As pointed out in ref.\cite{Johnson:2014yja}, since changing $\Lambda$ involves changing the $N$ of the dual theory, the heat engine cycle is a kind of tour on the {\it space} of field theories rather than staying within one particular field theory\footnote{As pointed out in ref.\cite{Karch:2015rpa} it also involves changing the size of the space the field theories live on.}. It is currently unclear as to the precise application of these heat engines in this context or any other, but as defined above they are  certainly   rich and well--defined enough in their own right to warrant an  exploration of their physics.

For now, focus on the cycle given in figure~\ref{fig:cyclesb}. In  ref.\cite{Johnson:2014yja},  it  is explained why such  a  choice is natural for  static black holes, which  we  study here\footnote{Refs.\cite{Belhaj:2015hha,Sadeghi:2015ksa,Caceres:2015vsa,Setare:2015yra,Johnson:2015ekr,Johnson:2015fva} have since done further studies of such heat engines.}.  The work performed is
$ W= \left(V_2-V_1\right)(p_1-p_4)$,  where the subscripts refer to the quantities evaluated at the corners labeled (1,2,3,4).  isochors are also adiabats for static black holes, and so the heat flows take place entirely  along the top and bottom, with the upper isobar  giving the  net inflow of heat and the lower isobar giving the exhaust:
 \begin{equation}
 Q_H=\int_{T_1}^{T_2} C_p(p_1,T) dT\ ,\qquad Q_C=\int_{T_4}^{T_3} C_p(p_4,T) dT\ ,
 \label{eq:hothothot}
\end{equation}
where  the  specific heat at constant pressure   $C_p\equiv T\partial S/\partial T|_{p}$. 

In the previous work on various black hole examples, the resulting efficiency $\eta=W/Q_H$ was evaluated  in a high temperature limit, since in general, the relations between $S$ and $T$ are such that the $T$--integral needed to evaluate $Q_H$ is difficult to perform exactly. Refs.\cite{Johnson:2015ekr,Johnson:2015fva} organize the computations of all needed quantities by  working in terms of the horizon radius $r_+$,  treating it as an independent parameter in terms of which all quantities can be most easily written. The high temperature expansions of all quantities were then  developed by working out the high temperature expansion of $r_+(T)$. It is worth noting here that  a next logical step could be to recast the $T$--integrals in equations~(\ref{eq:hothothot}) as  $r_+$--integrals. Since $C_p(r_+)$ is (for some classes of black hole) a ratio of polynomials in $r_+$, depending upon the nature of the $\partial r_+/\partial T$ Jacobian this might result in simpler expressions for the efficiency\footnote{This was also noted by Shao--Wen Wei, in a private communication.}. Indeed, as can be seen by evaluating a few examples, this  is correct. In fact, there's a much simpler way of looking at the whole picture that results in simple expression for the efficiency that makes it easy to see exactly why this works.  We will explore this next.

\section{A Simple Efficiency Formula}
The key point is that when black holes are the focus, it is the enthalpy $H$ that takes center stage in the First Law of Thermodynamics, since 
it is identified with the mass $M$ of the black hole\cite{Kastor:2009wy}. So instead of writing the First Law in terms of the internal energy, $dU=TdS-pdV$, one writes:
\begin{equation}
dH=TdS+Vdp\ ,
\end{equation}
and since our heat flows are along isobars, for which $dp=0$,  we see immediately that our total heat flow $\int TdS$, regardless of which variable we care to write it in terms of, is simply the {\it enthalpy change}, which is just the change in the black hole mass $M$. Therefore, we have a remarkably simple way to write our efficiency formula entirely in terms of the black hole mass evaluated at the corners:
\begin{equation}
\eta=1-\frac{M_3-M_4}{M_2-M_1}\ .
\label{eq:efficiency-prototype}
\end{equation}
The mass $M$ is most naturally written as a function of $r_+$ and $p$. Depending upon which parameters of the cycle are prescribed ({\it e.g.} scheme~1 $(T_1,T_2,p_1,p_4)$ or scheme~2 $(T_1,T_4,V_1,V_3)$ in refs.\cite{Johnson:2015ekr,Johnson:2015fva}), the equation of state can be readily used to evaluate the $r_{+,i}$ and hence the mass values. In fact, if the parameters specified are all pressures and volumes ({\it e.g.} $(p_1,p_4,V_1,V_2)$), the equation of state is not needed at all, (since the volume $V$ is a simple function  of $r_+$) and the resulting exact efficiency formula is remarkably  simple.

\subsection{Examples}
\label{sec:compare-high-T}
A nice class of examples is the static  black hole of charge $q$ in $D$ dimensions that has mass formula:
\begin{equation}
M=\frac{(D-2)\omega_{D-2}}{16\pi}\left(\alpha r_+^{D-5}+r_+^{D-3}+\frac{q^2}{r_+^{D-3}}+16\pi p \frac{r_+^{D-1}}{(D-1)(D-2)}\right)\ ,
\label{eq:GBmass}
\end{equation}
where for $\alpha=0$ we are in Einstein--Maxwell gravity, and for non--zero $\alpha$ we are in the Gauss--Bonnet generalization (see refs.\cite{Cai:2013qga,Johnson:2015ekr} for details and background). The relationship between $T,r_+$ and $p$ is given by:
\begin{equation}
T=\frac{1}{4\pi r_+(r_+^2+2\alpha)}\left(\frac{16\pi p r_+^4}{(D-2)}+(D-3)r_+^2+(D-5)\alpha-(D-3)\frac{q^2}{r_+^{2D-8}}\right)\ .
\label{eq:GBeqnofstate}
\end{equation}
 \begin{figure}[h]
{\centering
{\includegraphics[width=2.8in]{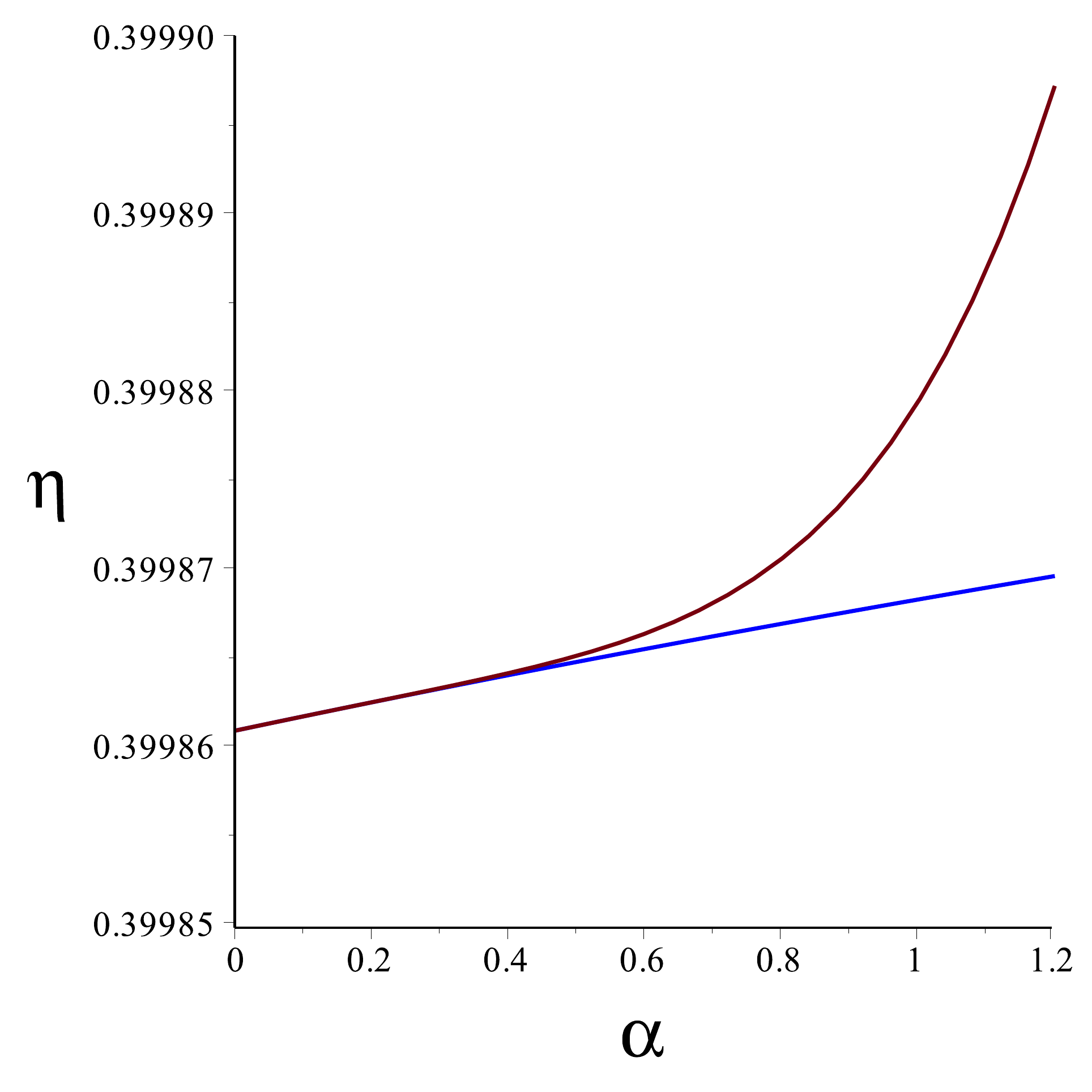} }
   \caption{\footnotesize   Comparison of the high temperature  result for the efficiency (upper curve), at fixed working temperatures, to the exact result (lower curve), as the Gauss--Bonnet parameter $\alpha$ is increased. Here we chose parameters $q=0.1,T_1=50,T_2=60, p_1=5$ and $p_4=3$.}  \label{fig:efficiency-comparison-alpha}
}
\end{figure}
There is no need to work at high temperature, as our formula~(\ref{eq:efficiency-prototype}) is exact. In fact, it is pleasing to see how the high temperature expansion is corrected by  it. In ref.\cite{Johnson:2015ekr}, for a given chosen (large) temperature, the expansions begin to break down once $\alpha$   gets too large.  Using the formula to compute the exact efficiency  is straightforward, and if (for example)  $T_1,T_2,p_1$ and $p_4$ are specified  parameters, one can invert equation~(\ref{eq:GBeqnofstate}) to compute the radii $r_{+,1}=r_{+,4}$ and $r_{+,2}=r_{+,3}$, and compare to the result for the large $T$ procedures done in ref.\cite{Johnson:2015ekr}.  Such an example is presented in figure~\ref{fig:efficiency-comparison-alpha}. As $\alpha$ increases, the deviation from accuracy (the upper curve is the large $T$ result) grows since the large $T$ expansion under--estimates the heat and over--estimates the work, as can be determined by a direct comparison of those quantities with the exact result.

\section{The Efficiency for Arbitrary Cycles}
\label{sec:arbitrary}
\begin{wrapfigure}{r}{0.4\textwidth}
{\centering
\includegraphics[width=2.8in]{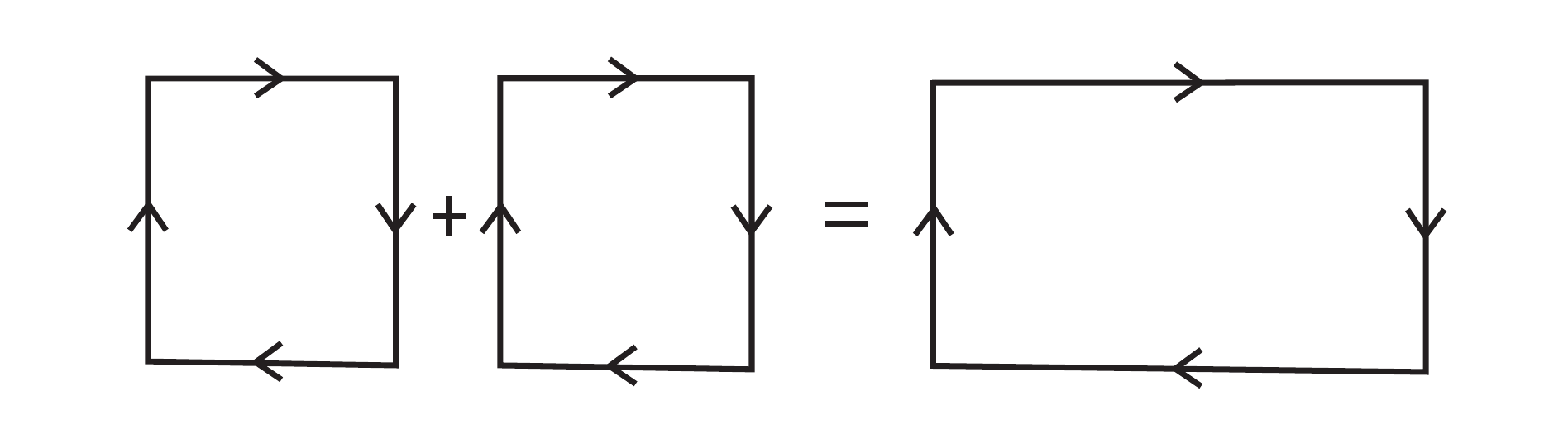} 
   \caption{\footnotesize  Adding cycles that share an edge.}   \label{fig:cyclesc}
}
\end{wrapfigure}

Working out the efficiency of more general shapes of cycle is in principle difficult, since the specific heat along the cycle is in general quite complicated, even for static black holes. Computing the heat flows would be quite a challenge.  A remarkable and immediate consequence of our result~(\ref{eq:efficiency-prototype}) is that it can be used as the basis for an algorithm for computing the efficiency of a cycle of {\it arbitrary shape} to any desired accuracy. Here's how: Cycles are additive, in the sense indicated in figure~\ref{fig:cyclesc}, and any closed shape in the $(p,V)$ plane can be approximated by tiling with   a regular lattice of rectangles.  
There will of course be parts where the tiling's edge  does not quite match the edge of the cycle's contour.  The amount of the failure can be reduced by simply shrinking the size of the unit cell.  At any given desired degree of accuracy (cell size), the efficiency is computed  simply as follows: Label all cells  at the corners as was  done for  the prototype cycle in figure~\ref{fig:cyclesb}. Only the cells at the edge contribute.  They do so by either having their upper edge open (no adjoining cell), in which case we call it a hot cell, or by having their lower edge open, a cold cell. Summing all the hot cell mass differences (evaluated at the top edges) will give~$Q_H$ and summing all the cold cell mass differences (evaluated at the bottom edges) will yield~$Q_C$, and the efficiency follows:
\begin{equation}
\eta=1-\frac{Q_C}{Q_H}\ , \qquad Q_H=\sum_{i{\rm th}\,\,{\rm hot}\,\, {\rm cell}}(M^{(i)}_2-M_1^{(i)})\ ,\qquad Q_C=\sum_{i{\rm th}\,\,{\rm cold}\,\,{\rm cell}}(M^{(i)}_3-M^{(i)}_4)
\label{eq:efficiency-master}
\end{equation}

\subsection{A Triangular Example}
\label{sec:triangular}
\begin{figure}[h]
{\centering
\subfigure[]{\includegraphics[width=2.8in]{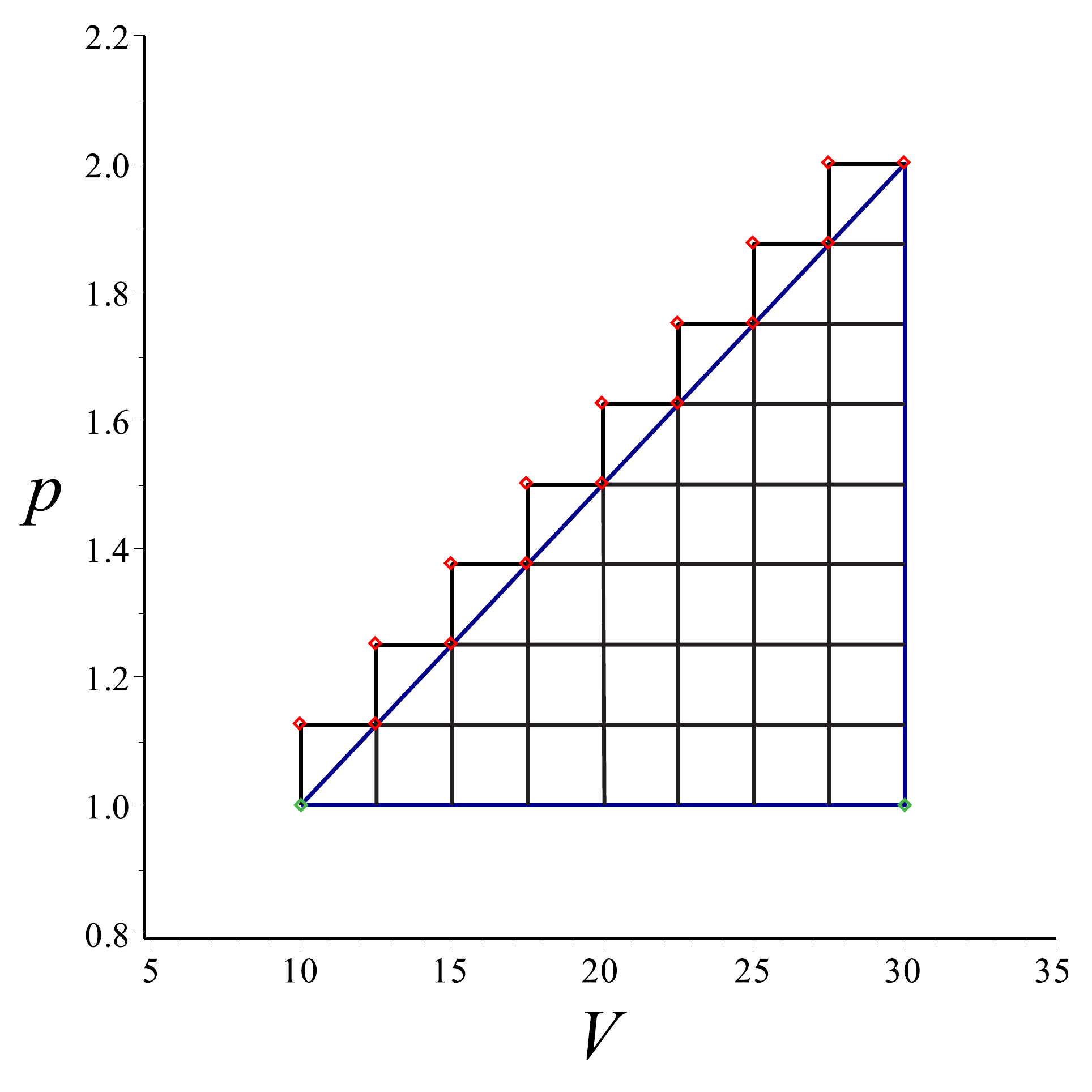} }\hspace{1.5cm}
\subfigure[]{\includegraphics[width=2.8in]{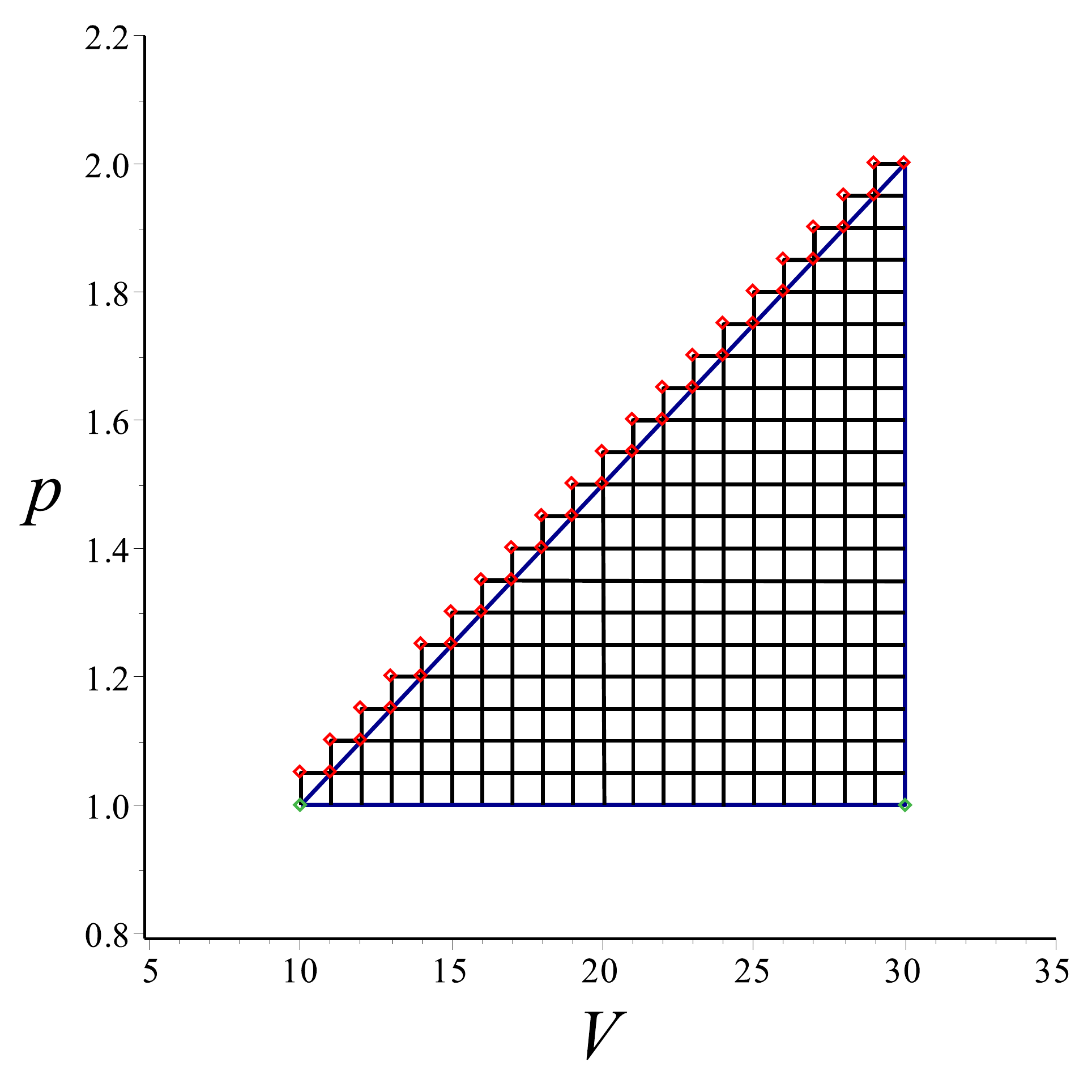}} 
   \caption{\footnotesize  (a) The discretized triangle with $N=8$.  (b) The discretized triangle with $N=20$.  The red dots mark the (upper) horizontal segments along which $Q_H$ is computed, and the green dots mark the (lower) horizontal segment along which $Q_C$ is computed. See text for discussion.}   \label{fig:discrete-triangle}
}
\end{figure}

\begin{figure}[h]
{\centering
{\includegraphics[width=2.8in]{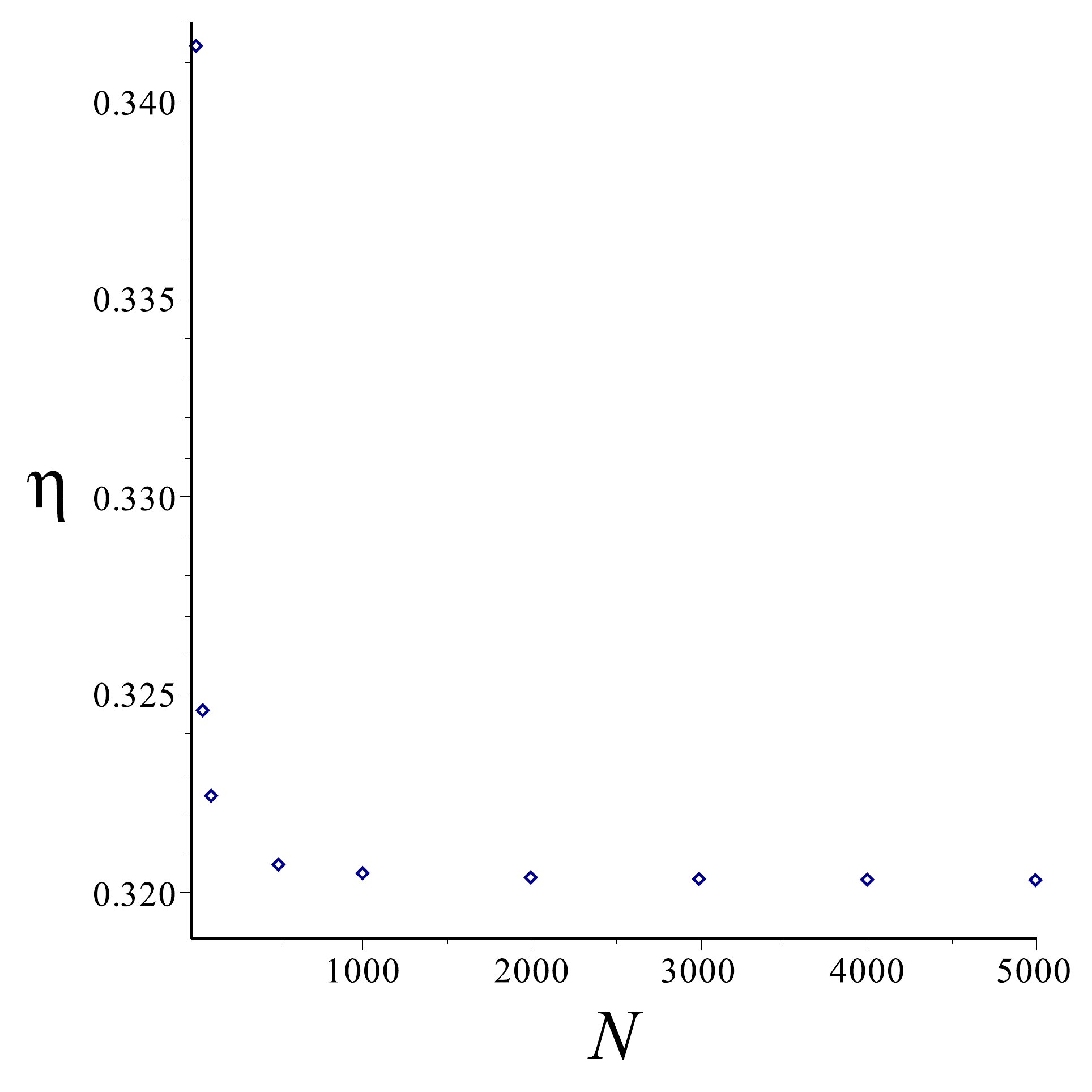} }
   \caption{\footnotesize  The convergence to the efficiency $\eta\sim0.32027$ as the discretizations of the triangle become    more accurate as $N$ increases. See text for discussion, and figure~\ref{fig:discrete-triangle}. Successive values include 0.34136, 0.32243, 0.32070, and 0.32030 for $N=10,100,500,$ and $5000$ respectively. }  \label{fig:efficiency_improvement}
}
\end{figure}

 A triangular  cycle was chosen to illustrate  the algorithm in action. It has two  edges parallel to the axes and one sloping edge in cycling from $(p,V)$ values $(2,30)$ to $(1,30)$ to $(1,10)$ and returning. One can choose to break it into a staircase--like rectangular lattice tiling that has $N$ rectangles (not to be confused with the $N$ of section~\ref{sec:introduction}) approximating the sloped edge, with better accuracy at larger $N$, as in figure~\ref{fig:discrete-triangle}. A simple  procedure was coded for the computation of the required $r_{+}$ and~$p$ values for each cell, summing the contributions to $Q_H$ according to our formula~(\ref{eq:efficiency-master}). In this case~$Q_C$ comes from the bottom edge  as the overall mass difference along the base.  Einstein--Maxwell black holes were used ({\it i.e.,} $\alpha=0$), choosing  the case $D=5$ in equation~(\ref{eq:GBmass}) for the mass. Figure~\ref{fig:efficiency_improvement} shows the rapid convergence toward the  result as $N$ is increased.

\section{Closing Remarks}
\label{sec:closing}
A remarkably simple expression~(\ref{eq:efficiency-master}) for the efficiency of a rectangular cycle  was presented, simplifying earlier presented expressions and allowing it to be computed exactly. The whole formula is in terms of the black hole mass evaluated at each of the four corners of the cycle. Since the mass is often readily computable in a closed form expression, this is an extremely compact result. Section~\ref{sec:compare-high-T} compared the results to the high temperature expansions obtained in earlier work. 

While it might seem relevant only to a special class of cycle, the expression turns out to be quite powerful, being the seed of an algorithm for computing the efficiency of cycles of arbitrary shape using a well--defined geometrical procedure. Such shapes would in general be hard to compute the efficiency for by {\it e.g.} numerical integration methods along the path. The presented example (the triangle of section~\ref{sec:triangular}) shows how rapidly the algorithm  converges to a result for the efficiency.

In fact, the procedure extends to wider classes of heat engine, whether a black hole is used as a working substance or not. The key object to be able to compute with is the basic unit cell needed to tile a cycle of arbitrary shape. More generally, the unit cell is actually a Brayton/Joule cycle, composed of two isobars and two adiabats. (In this paper it is a rectangle, since for static black holes,   adiabats are isochors). So all the heat flows are taken into account on the top and bottom edges.  Since  they are  isobars,  the heat flows can be written entirely in terms of the enthalpy. So for any system for which one can readily compute  the enthalpy $H$ (and black holes turn out to be such a system since  $H$ is simply the mass $M$), and for which one knows the adiabatic curves, the same  geometrical algorithm can be used to compute the efficiency of an arbitrary cycle in the manner described in section~\ref{sec:arbitrary}. It would be interesting to study some examples of this application.

\bigskip

\section*{Acknowledgements}
 CVJ would like to thank the  US Department of Energy for support under grant DE-FG03-84ER-40168,  and Amelia for her support and patience.


\providecommand{\href}[2]{#2}\begingroup\raggedright\endgroup

\end{document}